%% file: main.tex
\documentclass[sigconf,10pt,nonacm]{acmart}
\usepackage{graphicx}
\usepackage{enumitem}
\usepackage{tikz}
\usepackage{adjustbox}
\usepackage{subfigure}
\usepackage{xcolor}
\usepackage{circledsteps}
\usepackage{enumitem}

\newcommand\myCircled[2][]{\ifmmode
\Circled[fill color=black,inner color=white,#1]{\mathsf{#2}}
\else
\Circled[fill color=black,inner color=white,#1]{\sffamily#2}
\fi}

\title{Rethinking Image Compression on the Web with Generative AI}

\author{Shayan Ali Hassan, Danish Humair, Ihsan Ayyub Qazi, Zafar Ayyub Qazi}
\affiliation{%
  \institution{Department of Computer Science\\Lahore University of Management Sciences (LUMS)}
  \city{Lahore}
  \country{Pakistan}
}
\email{{25100165, 25100183, ihsan.qazi, zafar.qazi}@lums.edu.pk}

\input{abstract}

\begin{document}

\maketitle

\input{introduction}

\input{motivation}

\input{framework}

\input{methodology}

\input{userstudy}

\input{discussion}

\bibliographystyle{ACM-Reference-Format} 
\bibliography{main}

\end{document}

%% file: abstract.tex
\begin{abstract}
The rapid growth of the Internet, driven by social media, web browsing, and video streaming, has made images central to the Web experience, resulting in significant data transfer and increased webpage sizes. Traditional image compression methods, while reducing bandwidth, often degrade image quality. This paper explores a novel approach using generative AI to reconstruct images at the edge or client-side. We develop a framework that leverages text prompts and provides additional conditioning inputs like Canny edges and color palettes to a text-to-image model, achieving up to 99.8\% bandwidth savings in the best cases and 92.6\% on average, while maintaining high perceptual similarity. Empirical analysis and a user study show that our method preserves image meaning and structure more effectively than traditional compression methods, offering a promising solution for reducing bandwidth usage and improving Internet affordability with minimal degradation in image quality.
\end{abstract}

%% file: introduction.tex
\section{Introduction}
\label{sec:introduction}
The rapid growth of the Internet, driven largely by web browsing, social media, and video streaming, has made images central to our Web experience. This expansion has resulted in a staggering amount of data being transferred daily, with approximately 321\,Exabytes transmitted each day. Web browsing and social media account for 18.4\% of this traffic, while video streaming accounts for 53.7\%~\cite{sandvine2023global}. Consequently, a significant number of images are transmitted over the Internet as thumbnails, social media posts, and other website elements. Images typically contribute the most bytes to webpages, accounting for 24\% to 58\% of the total size. Additionally, we are observing rapidly increasing webpage sizes, driven partly by the inclusion of more and higher-quality images, leading to a 13-fold increase in median webpage sizes over the last decade~\cite{habib2023framework}.

This growth has significant implications for Internet affordability and inclusion, as well as ingress bandwidth requirements for ISPs and CDNs.
According to the UN Broadband Commission for Sustainable Development, 94 developing countries fail to meet the target for affordable broadband services, which states that Internet plan costs should not exceed 2\% of the monthly Gross National Income (GNI) per capita \cite{itu}.
Thus, reducing webpage sizes is crucial to making the Internet more affordable and equitable for people in these countries. Additionally, most Internet traffic originates from large data centers and is delivered to edge computing clusters, such as Content Delivery Networks, which are closer to end users. Consequently, saving even a small fraction of the data transferred between these entities can significantly reduce ingress bandwidth requirements and the overall cost of sending content over the network. Simultaneously, reducing these bandwidth requirements may also impact data egress charges from cloud providers for moving or transferring data from the cloud storage where it was uploaded~\cite{kaufmann2015price}.%

To reduce the data footprint of images, image compression is widely used on the Web. Traditional lossless image compression formats, such as PNG, can achieve compression ratios of up to 3:1 (66\% bandwidth reduction), while lossy formats, such as JPEG and WebP, can achieve ratios up to 20:1 (95\% bandwidth reduction) \cite{pintus2012objective}. However, at these high compression levels, image quality often degrades significantly due to compression and noise artifacts, negatively impacting the user browsing experience. We explore a radically different approach to image compression by reconstructing images at the edge or client-side through generative AI.

The rapid evolution of generative AI, particularly in the realm of image synthesis and reconstruction, opens up exciting possibilities for image optimization. By leveraging advanced text to image generative models like Stable Diffusion, we can potentially achieve significant bandwidth savings while maintaining or even enhancing image quality. In this paper, we explore a design space of using generative AI in image optimization, ranging from text-only prompts to more sophisticated approaches that involve passing additional conditioning inputs to a generative model to improve accuracy and control. We discuss challenges related to ensuring semantic accuracy, particularly for critical elements like faces, text, and logos, minimizing hallucinations and unwanted artifacts, balancing compression ratios with reconstruction quality, and optimizing generation speed for real-time applications

To make our exploration concrete, we develop a framework for reconstructing source images at the edge using Stable Diffusion, thus bypassing the need for transmitting images entirely. Our framework utilizes not only text prompts but also investigates the use of additional conditional inputs such as Canny edges~\cite{canny1986computational} and color palettes. We also provide an in-painting procedure for preserving perceptually important details and ensuring semantic accuracy. Overall, through our design space exploration and framework, we shed light on the following key research questions:
\begin{enumerate}[leftmargin=*,topsep=0pt,label=\protect\myCircled{\arabic*}]
    \setlength{\itemsep}{0pt}
    \setlength{\parskip}{0pt}
\item \textit{Can generative AI be used to faithfully reconstruct source content at the edge? If so, which guiding factors can be sent to maximize similarity with the original image?}
\item \textit{How can we measure the similarity of recreated images with the originals? How do different guiding factors affect this similarity?}
\item \textit{How do we quantify the bandwidth–similarity trade-off? In what contexts can we use source content reconstruction, and how does it align with human ratings?}
\end{enumerate}

Based on our empirical analysis of a dataset comprising 490 images from the top 1000 globally most visited websites and a synthetic dataset of high-quality images, we observe image bandwidth savings of up to 99.8\% in the best case (with text prompts only) and 92.6\% on average (with a few additional conditioning inputs). The VGG16 perceptual similarity metric \cite{simonyan2014very} indicates a high degree of preservation of image meaning and structure, with average scores of 0.82 for the most accurate reconstruction strategy and 0.76 on average (normalized between 0 and 1) (\S\ref{sec:experiments}). These findings are supported by a user study involving 22 participants, where respondents rated the reconstructed images from our framework as more structurally similar to the original images roughly half the time compared to traditionally compressed images. Additionally, respondents rated our reconstructed images as accurately preserving the overall image meaning. Notably, in 5 out of 8 cases, respondents preferred viewing our reconstructed images over compressed images, despite both requiring the same bandwidth for transmission (\S\ref{sec:userstudy}).

%% file: motivation.tex
\section{Motivation}
\label{sec:motivation}
Image compression techniques are crucial for efficient data transmission, storage, and Internet affordability but they often involve a trade-off between file size and visual quality~\cite{wang2004image,habib2023framework}. Recent advances in generative AI, including the development of the transformer architecture \cite{vaswani2017attention} and denoising diffusion probabilistic models (DDPMs) \cite{ho2020denoising}, along with significant improvements in Large Language Models (LLMs) such as GPT-4 and prompt-based image synthesis models, have led to the increasing popularity of AI-based image compression formats and upscaling methods.
Furthermore, recent algorithmic image compression formats, such as WebP, reportedly offer 39.8\% greater byte-size efficiency compared to JPEG at the same quality level \cite{ginesu2012objective}. However, concerns remain with both approaches.

\subsection{Limitations of Existing Compression Approaches}
At high compression levels, image quality degrades significantly due to compression artifacts and noise, adversely affecting users' browsing experience. High compression typically leads to noticeable degradation in image quality. The Structural Similarity Index Measure (SSIM), a common metric for assessing perceptible loss, indicates that SSIM values below 0.8 denote visible artifacts and degradation \cite{wang2004image, wang2009mean}. We collected 150 high-resolution JPEG and WebP images from popular websites, reduced their quality to achieve a 90\% reduction in file size, and measured the perceptible loss using SSIM. We observed an average SSIM value of 0.77, aligning with prior studies on the impact of high compression on image quality.

AI-based super-resolution, which upscales low-resolution images, can mitigate quality loss from compression. However, upscaled images often lack intrinsic noise that provides finer texture details, resulting in perceptually unsatisfying views \cite{ledig2017photo}.
This occurs because compression irreversibly removes certain details, making accurate retrieval impossible without additional information from the original image.
This is particularly problematic for small details that humans are highly perceptive to or those that have critical implications if inaccurately displayed, such as human faces, text, and brand logos (which are essential for brand identity). Traditional compression methods risk obscuring critical details that AI-based upscaling may not fully recover.
This underscores the need for specialized approaches balancing compression efficiency with preservation of semantically important features.

\subsection{Opportunities and Challenges with GenAI: A Design Space Exploration}

Recent advances in generative AI, particularly in image reconstruction, offer promising opportunities for image optimization. For example, Stable Diffusion, a latent text-to-image DDPM, achieves state-of-the-art performance in generating high-quality, photo-realistic or stylized images from textual prompts. Reconstructing images with generative AI offers three key benefits:

\begin{itemize}[leftmargin=*,topsep=0pt]
    \setlength{\itemsep}{0pt}
    \setlength{\parskip}{0pt}
    \item \emph{Enhanced Compression Ratios.} Generative AI allows for the transmission of `conditioning inputs' instead of the full image, potentially achieving higher compression ratios without sacrificing image quality.
    \item \emph{Mitigation of Compression Artifacts.} By generating images from prompts, generative AI can mitigate artifacts typically introduced by traditional compression methods. Furthermore, images can be natively generated at various resolutions regardless of original image size, providing high quality and fidelity even at lower resolutions.
    \item \emph{Preservation of Perceptually Important Details.} Generative models can specifically focus on preserving details which are often degraded by traditional compression but crucial to human perception, such as faces, text, and logos.
\end{itemize}

\begin{figure}
    \centering
    \includegraphics[width=0.45\textwidth]{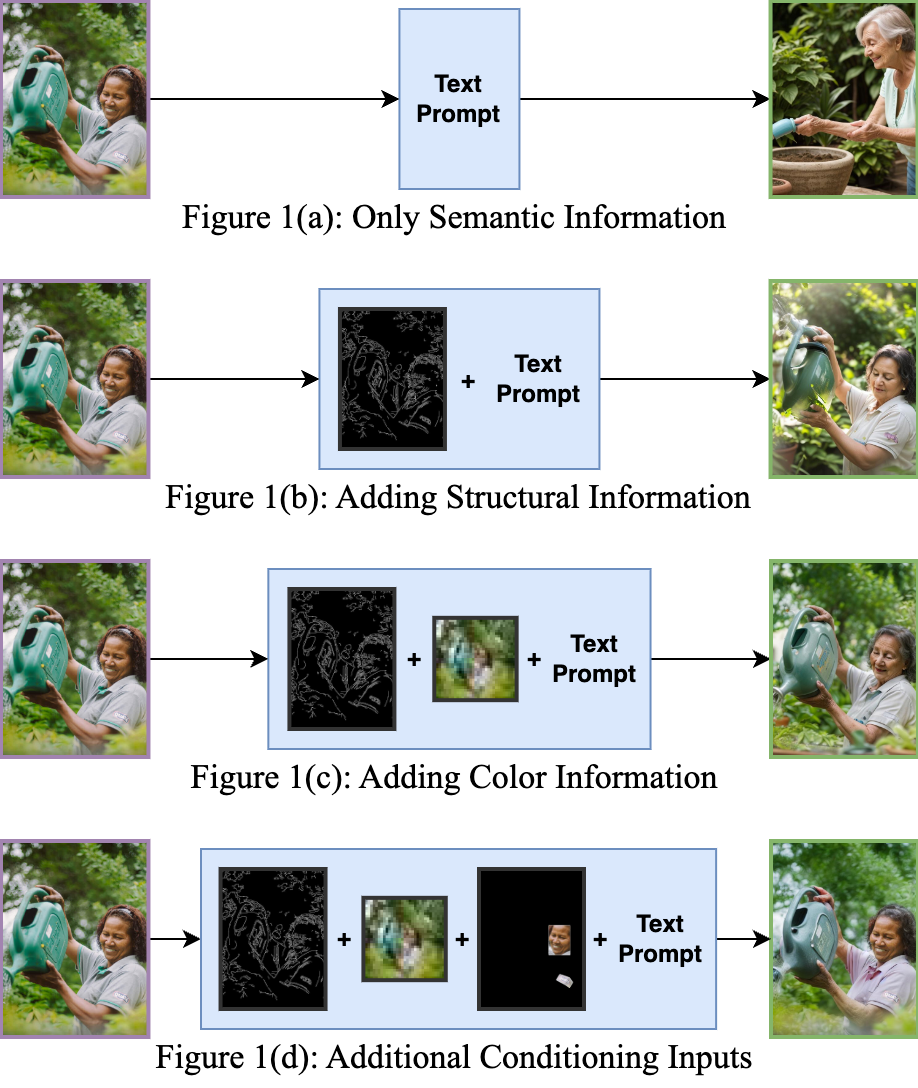}
    \vspace{-0.15in}
    \caption{Design space of using generative AI for image optimization.}
    \vspace{-0.15in}
    \label{fig:design_space}
\end{figure}

We now explore a design space of using generative AI for image optimization, ranging from text-only prompts to more sophisticated approaches:

\noindent\textbf{1. Text-Only Prompts for Image Generation:} At its most basic level, we can use text prompts to generate images from scratch. Given that the standard Unicode encoding form is 16-bit, and that most images on the internet are on the order of kilobytes, such an approach can provide unprecedented bandwidth savings. However, while the holistic, semantic information can be conveyed in this manner, ensuring accurate recreation of image composition and structure remains challenging; see Figure \ref{fig:design_space}. This method could thus be particularly useful for placeholder images, illustrations based on textual content, stock images and images that aim to illustrate broad concepts.

\noindent\textbf{2. Edge Detection + Text Prompt:}
We can additionally use edge detection techniques (e.g., the Canny edge detection algorithm) to provide more structural guidance to generative AI models as shown in Figure \ref{fig:design_space}. This approach strikes a balance between minimal data transmission and improved accuracy, as edge information is typically compact. The detected edges help maintain the overall composition and key features of the original image, while the text prompt guides the model in filling in details and style.

\noindent\textbf{3. Color Palettes + Edge Detection + Text Prompt:} Downsampling the image to a low-resolution color grid (e.g., 32x32) provides color information to the generative model while maintaining significant bandwidth reduction. This color data allows for accurate color reproduction in each structurally distinct part of the reconstructed image, enhancing fidelity to the original.

\noindent\textbf{4. Additional Conditioning Inputs + Text Prompt:} To further enhance accuracy and control and allow for more precise and faithful image reconstruction, we can incorporate additional conditioning inputs alongside text prompts. These could include semantic segmentation masks, low-resolution thumbnails, style transfer parameters and normal maps to name a few.

As we delve into this design space, few key challenges present new opportunities. There needs to be a way to accurately preserve critical elements like faces, text, and logos, while also minimizing hallucinations and unwanted artifacts. There is also a need to optimize generation speed for real-time applications, as well as balancing compression with reconstruction quality.

%% file: framework.tex
\section{Pseudo-Lossy Compression Framework}
\label{sec:framework}

\begin{figure}[t]
    \centering
    \includegraphics[width=\columnwidth]{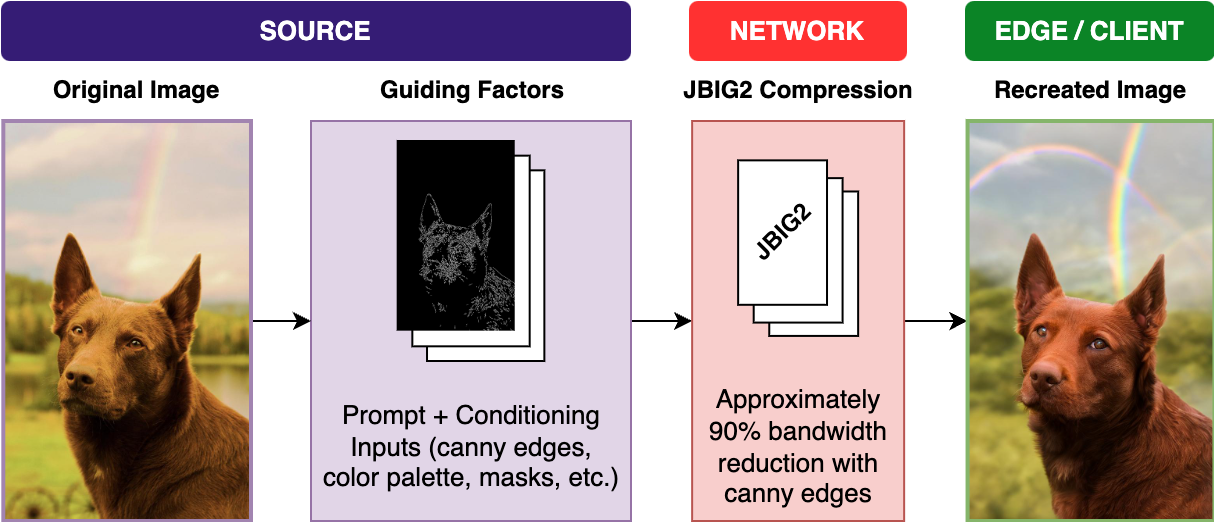}
    \caption{The Pseudo-Lossy Compression Framework.}
    \vspace{-0.25in}
    \label{fig: The Pseudo-Lossy Compression Framework}
\end{figure}

Traditional compression algorithms cannot perceptually identify the information that has the greatest impact on our viewing experience, they simply opt to reduce the precision of high-frequency details, leading to perceptually unsatisfying images. In contrast, we propose extracting only the semantic and structural information as \textit{conditioning inputs} from the original image, a technique which can be considered as `Pseudo-Lossy' Compression. By extracting such information from the original image itself, we guarantee it's preservation throughout the reconstruction process, unlike upscaling where finer details lost due to down-sampling are impossible to perfectly reconstruct back without access to some information from the original image again.

In this work, we recommend four conditioning inputs as part of our Pseudo-Lossy Compression framework which aim to maximize perceptual similarity and bandwidth savings:

\noindent\textbf{1. Text prompt:} The prompt serves as a description of the image and is sufficient to provide semantic information.

\noindent \textbf{2. Canny Edges:} The canny edges provide the structural information, ensuring that reconstructions have the same composition as the original.

\noindent \textbf{3. Color Palettes:} A color palette further refines the structural information provided to the generative model by ensuring accurate color replications.

\noindent\textbf{4. In-painting Masks:} Stable Diffusion Inpainting involves using a black and white mask to indicate which sections of an image should be recreated, and which sections should be left untouched, thus allowing for the preservation of critical, \textit{salient features}, such as faces, small text and logos. Keeping the masks rectangular allows them to represented by a series of coordinates, further saving bandwidth. The cropped salient feature would need to be transmitted unmodified, hence the total bandwidth savings in this case would be indirectly proportional to the area of salient features in the original image. However, since only smaller important details are at risk of being distorted, bandwidth savings are expected to still be significant.

A high-level overview of the framework is shown in Figure \ref{fig: The Pseudo-Lossy Compression Framework}. We also present further bandwidth optimizations as well as a method for evaluating the effectiveness of our framework.

\subsection{Additional Optimizations}
\begin{figure}[t]
    \centering
    \includegraphics[width=\columnwidth]{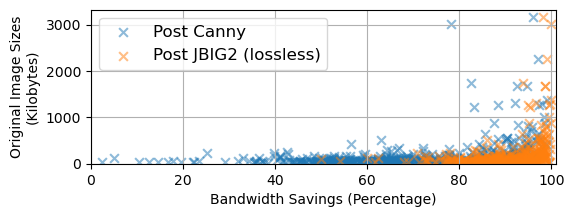}
    \vspace{-0.15in}
    \caption{Most bandwidth savings come from converting images to Canny edge-maps and further reductions can be obtained by utilizing JBIG2 compression.}
    \vspace{-0.1in}
    \label{fig: Canny / JBIG2 savings}
\end{figure}

As previously discussed, the original image can be compressed significantly using Canny edge detection, up to a factor of 8, because the edges are monochromatic. Due to edge detection, most pixels in a Canny bitmap are the same color, representing negative space between thin, edge outlines, allowing further optimization.
We exploit this by using lossless JBIG2, the industry standard compression algorithm for bi-level images, released in 2000 for fax machines and black-and-white PDFs. JBIG2 outperforms other algorithms by reducing bi-level images by a factor of 2-5 \cite{howard1998emerging}. In our representative dataset of 490 images (\S\ref{sec:experiments}), we observed a strong positive correlation between original image size and bandwidth savings, as shown in Figure \ref{fig: Canny / JBIG2 savings}. JBIG2 provides significant bandwidth reduction, up to 99.95\% in the best case and around 90\% on average in our experiments.

\subsection{Measuring Perceptual Similarity}
A robust metric for comparing the semantic and structural similarity of images pre and post reconstruction is extremely important, as that allows us to explore the bandwidth--similarity trade-off when reconstructing images over a network. To measure perceptual similarity, we cannot rely on pixel-level metrics such as SSIM or PSNR. Even if the two images are structurally and semantically the same to a human, at the pixel level they are significantly different due to Stable Diffusion 'filling in' the Canny edges with what it thinks is appropriate. Thus the intrinsic noise that adds finer details is different between the original and the reconstruction.

We utilise the VGG16 convolutional neural network (CNN) for quantifying perceptual similarity \cite{simonyan2014very}. By using VGG16, we can extract high-level, holistic features from two images and compare them. The model creates an embedding vector for each image and and we can calculate the cosine similarity between them, with 0 indicating dissimilarity and 1 indicating similarity. For later reference, we calculate VGG scores between 2 random and perceptually dissimilar images, and repeat the process 50 times. The average score was 0.32.

%% file: methodology.tex
\section{Methodology and Experiments}
\label{sec:experiments}

In this section, we will describe the experimental methodology for evaluating our framework, including the collection process for our representative and synthetic datasets, the experiments conducted and our results.

\subsection{Dataset Collection}

\textbf{Representative Dataset}. 
We scraped images from the top 1000 globally visited domains using the Google Chrome UX Report (CRUX) lists \cite{ruth2022toppling}. We manually removed pages with inappropriate content resulting in a final set of 600 pages. Given the imbalance in image quantity across websites, each website was categorized into one of seven main categories derived from Cloudflare's domain categorizations \cite{ruth2022world}: E-commerce, Informational, Business/Company, News, Social Media, Video Streaming, and Other. 

From each category, we randomly sampled 70 images, resulting in a dataset of 490 images. Images with transparent pixels were purposefully excluded due to incompatibility with the models being used, and a minimum resolution of $512\times256$ in either orientation was also set, as images smaller than this resolution would neither work well with text-to-image models nor even benefit as much from major bandwidth savings as they were already small. The set of images collected ranged from dimensions $333\times687$ to $4032\times2030$ and included file types PNG, JPEG and WebP.

\noindent\textbf{Synthetic Dataset}. 
Our framework produces high-quality images with significant data savings. However, many images on the top 1000 globally visited domains are already compressed and of reduced quality. To better demonstrate our framework's potential, we curated a synthetic dataset. We collected 150 high-resolution images from news websites such as BBC News and The New York Times, as well as Wikipedia, where images are consistently high-quality and diverse in context.

\begin{figure*}[ht]
    \centering
    \begin{subfigure}
        \centering
        \includegraphics[width=5.7cm]{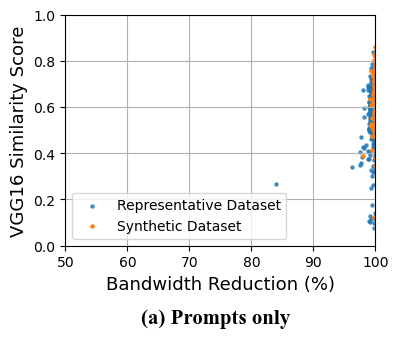}
    \end{subfigure}%
    \begin{subfigure}
        \centering
        \includegraphics[width=5.7cm]{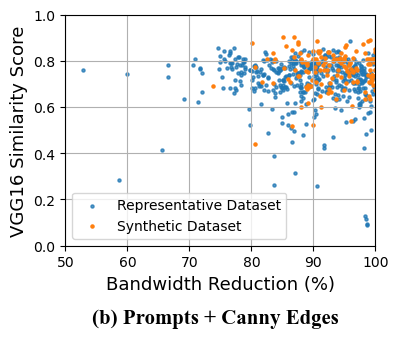}
    \end{subfigure}%
    \begin{subfigure}
        \centering
        \includegraphics[width=5.95cm]{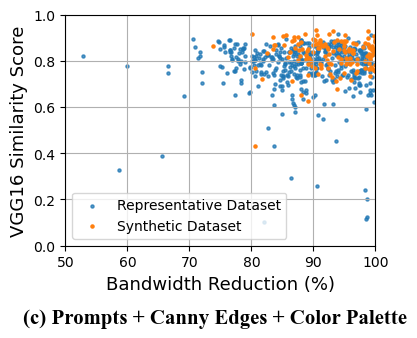}
    \end{subfigure}%
    \vspace{-0.15in}
    \caption{This figure illustrates the trade-off between expected bandwidth reduction and similarity (measured by VGG16) for the first three experiments.}
      \vspace{-0.15in}
    \label{fig: experimental_results}
\end{figure*}
\subsection{Experiments}

The main goal of our experimental methodology was to assess the bandwidth reduction---similarity trade-off and explore how it can be affected by different combinations of conditioning inputs. Thus we conduct four experiments:

\noindent\textit{Prompts Only:} In the baseline experiment, only text prompts were used to generate the images using a fine-tuned version of the Stable Diffusion 1.5 text-to-image model. These prompts were generated by sending zero-shot instructions and the original image to the GPT-4o API. %

\noindent\textit{Prompts + Canny Edges:} In the next experiment, structural information was also added as a conditioning input in the form of canny edges. A separate ControlNet model was used to condition the reconstruction to fit within these structural constraints, and the canny edges were compressed with JBIG2.

\noindent \textit{Prompts + Canny Edges + Color Palette:} In the third experiment, we add a WebP compressed color palette to provide color specific structural information. %

\noindent \textit{Salient Features:} In many cases, small details of interest such as faces, logos, and small text, were too small to be accurately defined by canny edges, and the color information was also not enough to accurately reconstruct them. For this experiment, we randomly sampled 50 such images and manually selected these salient features, and used two ControlNet pipelines to condition the input, where one conditioned the image generation based on structure and color and the other used inpainting to generate all parts of the image other than the selected region. This region was also considered in expected bandwidth usage.

\subsection{Results}

Our experimental results are shown in Figure \ref{fig: experimental_results} and Table \ref{tab:experiments_table}. \footnote{The negative bandwidth savings instance in the salient features experiment is probably due to the original image being extremely small, thus the conditioning inputs were larger than the image itself. In an later implementation of this framework, checks will be in place to prevent this from happening.} For the prompt-only experiment, we observe the highest bandwidth savings but also the lowest VGG scores. However, the scores are still significantly higher than comparing two completely dissimilar pair of images ~\S\ref{sec:framework}. For the Canny experiment, we observe high bandwidth savings and higher VGG scores, with slightly more savings in our synthetic dataset. By adding color palettes, bandwidth savings remain constant however VGG scores become even higher for both datasets. For salient feature preservation, we observe the highest median similarity score, indicating the utmost structural and semantic preservation. However, this comes at the cost of the relatively lowest bandwidth savings, since the salient features must be transmitted unmodified.

\begin{table}
\centering
\resizebox{0.47\textwidth}{!}{%
\begin{tabular}{llcccccc}
\toprule
\textbf{Experiment} & \textbf{Result} & \multicolumn{3}{c}{\textbf{Synthetic Dataset}} & \multicolumn{3}{c}{\textbf{Representative Dataset}} \\
\cmidrule(r){3-5} \cmidrule(r){6-8}
 & & \textbf{Median} & Min & Max & \textbf{Median} & Min & Max \\
\midrule
\textbf{Prompt only} & Bandwidth Savings & \textbf{99.93} & 97.95 & 99.99 & \textbf{99.70} & 84.00 & 99.99 \\
& VGG Scores & \textbf{0.64} & 0.13 & 0.86 & \textbf{0.60} & 0.07 & 0.84 \\
\midrule
\textbf{Prompt +} & Bandwidth Savings & \textbf{92.92} & 73.75 & 99.94 & \textbf{89.39} & 52.91 & 99.86 \\
\textbf{Canny} & VGG Scores & \textbf{0.76} & 0.44 & 0.90 & \textbf{0.72} & 0.09 & 0.88 \\
\midrule
\textbf{Prompt +} & Bandwidth Savings & \textbf{92.92} & 73.75 & 99.94 & \textbf{89.39} & 52.91 & 99.86 \\
\textbf{Canny + Color} & VGG Scores & \textbf{0.82} & 0.43 & 0.93 & \textbf{0.77} & 0.10 & 0.91 \\
\midrule
\textbf{Salient} & Bandwidth Savings & & & & \textbf{77.85} & -81.52 & 96.82 \\
\textbf{Features} & VGG Scores & & & & \textbf{0.86} & 0.65 & 0.93 \\
\bottomrule
\end{tabular}}
\caption{Results of various experiments with similarity scores and bandwidth savings.}
\label{tab:experiments_table}
\vspace{-0.25in}
\end{table}

%% file: userstudy.tex
\section{User Study}
\label{sec:userstudy}

\begin{figure*}[t!]
    \centering
    \begin{subfigure}
        \centering
        \includegraphics[width=5.7cm]{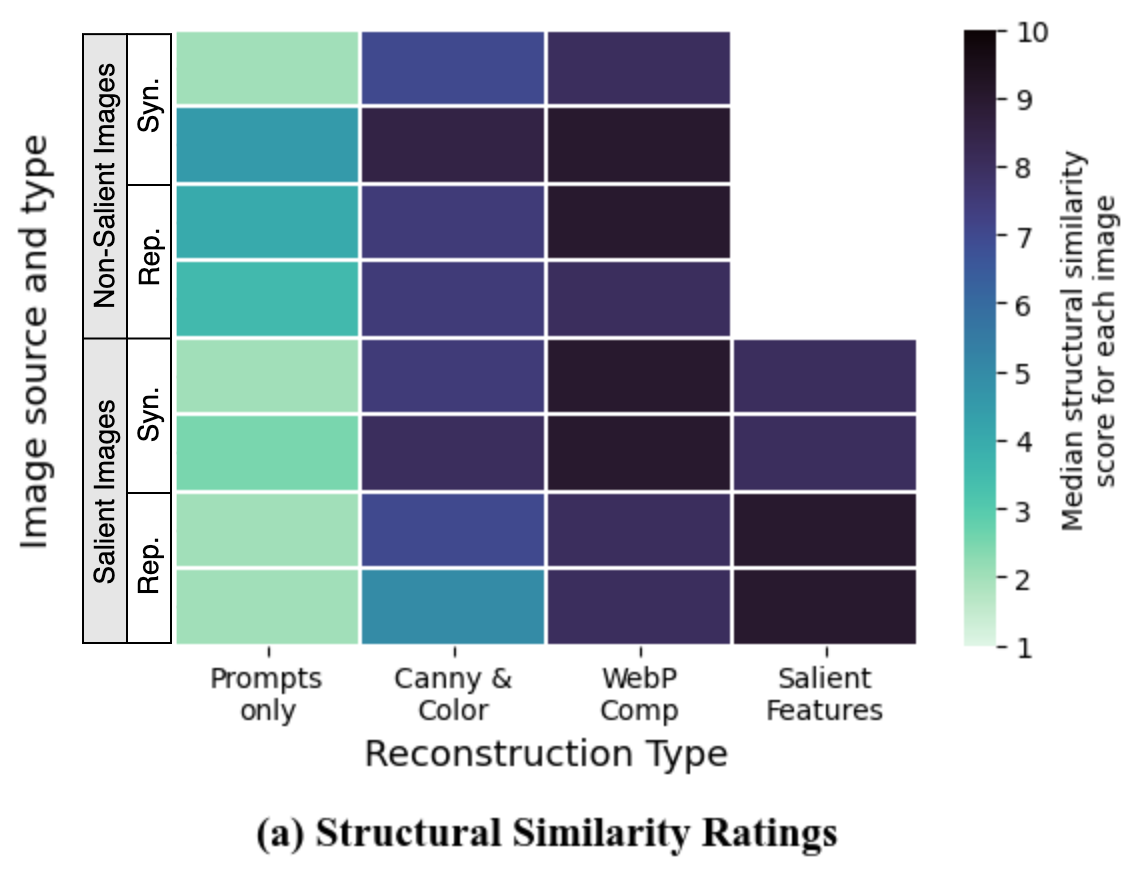}
    \end{subfigure}%
    \begin{subfigure}
        \centering
        \includegraphics[width=5.7cm]{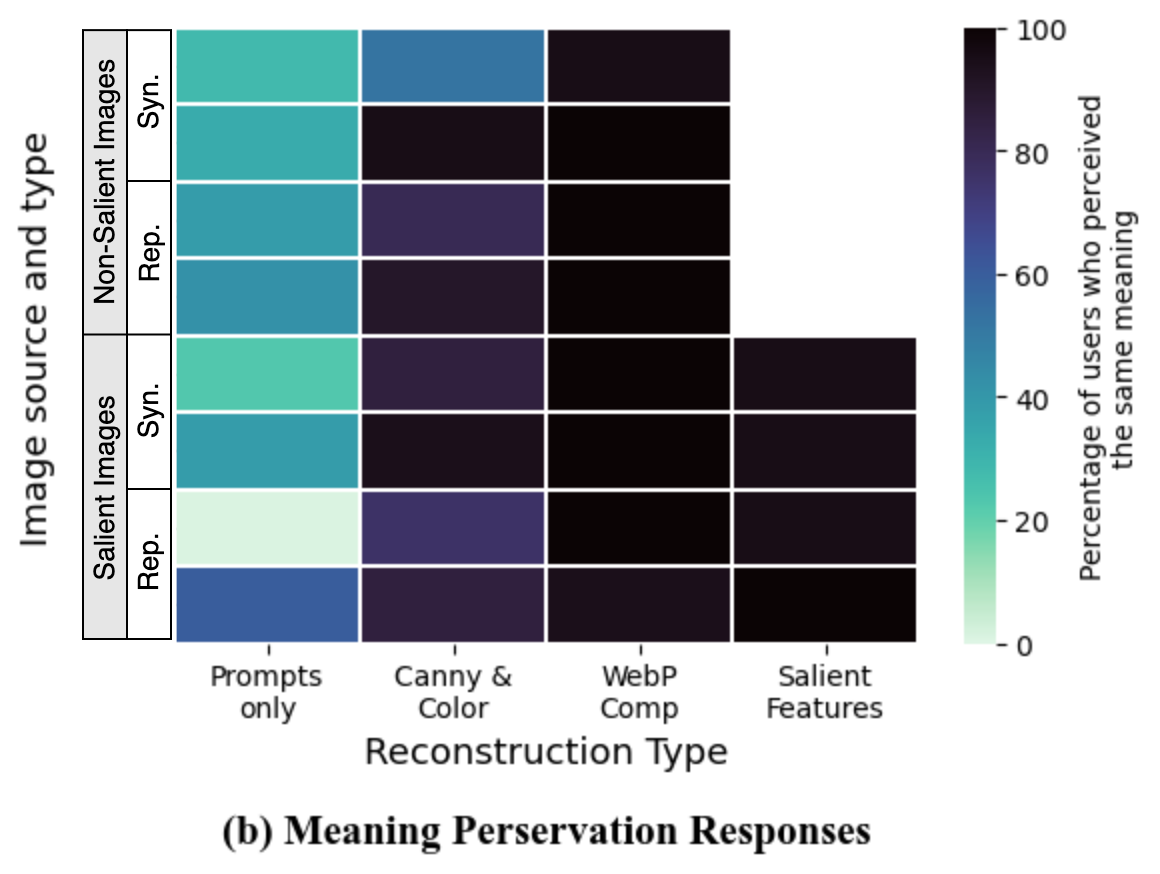}
    \end{subfigure}%
    \begin{subfigure}
        \centering
        \includegraphics[width=5.3cm]{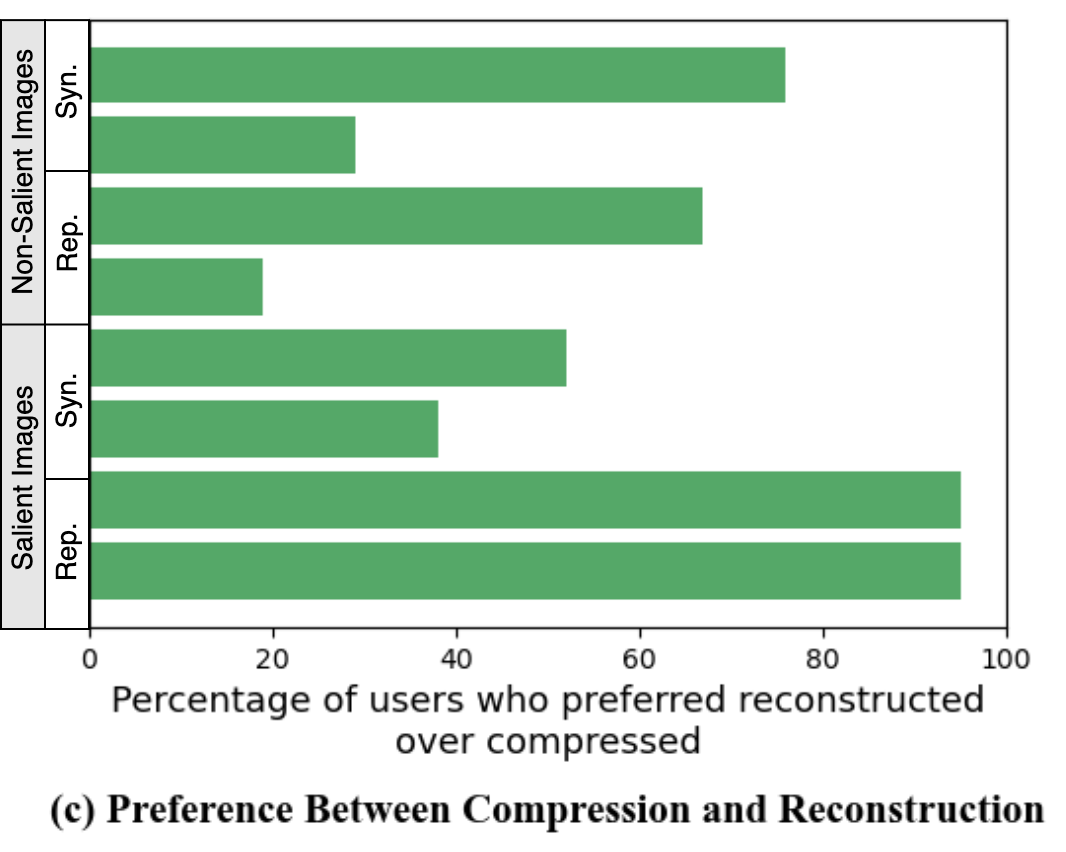}
    \end{subfigure}%
    \vspace{-0.15in}
    \caption{Results from the user study showing (a) structural similarity ratings, (b) meaning preservation responses, and (c) preference between compressed and reconstructed images.}
    \label{fig: user_study_responses}
    \vspace{-0.15in}
\end{figure*}

To evaluate the effectiveness of our framework through human perception, we conducted a user study. We randomly selected 8 images, evenly divided into two categories: 4 with salient features to preserve and 4 without. To ensure diversity, we drew half of the images from our representative dataset and half from the synthetic dataset for each category. The study aimed to examine how different conditioning inputs influence image perception. For each image, we included the following versions in our assessment:

\begin{itemize}[leftmargin=*,topsep=0pt]
    \setlength{\itemsep}{1pt}
    \setlength{\parskip}{0pt}
    \setlength{\parsep}{0pt}
    \item \textbf{Original:} The unaltered image at its original resolution.
    \item \textbf{Prompt Only:} Image reconstructed using only GPT-4 prompts, preserving semantic information.
    \item \textbf{Canny and Color:} Image reconstructed using prompts, canny edges, and color palettes, retaining semantic and structural information.
    \item \textbf{Salient:} When applicable, image reconstructed with prompts, canny edges, color palettes, and salient feature masks, preserving semantic, structural, and salient information.
    \item \textbf{WebP Compressed:} Image compressed to match the bandwidth of Salient or Non-Salient versions, using WebP for optimal quality at equivalent file sizes.
\end{itemize}

The user study was divided into three main sections. In Section A, respondents were shown the original image alongside all other versions of each image and rated the structural similarity on a scale of 1 to 10. Section B involved showing respondents the original image within its web-page context, and for all other versions, they were asked if replacing the original image would preserve the web-page's intended meaning. In Section C, respondents were presented with the compressed and Salient reconstruction (or Canny and Color if no salient reconstruction was available) and asked to indicate which one they preferred.

The results of the user study are shown in Figure \ref{fig: user_study_responses}. Half of the respondents rated the salient images as more structurally similar to the originals than the compressed versions, with Canny and Color images also receiving high ratings. Additionally, salient images were considered equally effective in preserving the webpage's intended meaning. Notably, for 5 out of 8 images, respondents preferred our reconstructed images over the compressed ones, despite both requiring the same bandwidth. This preference likely stems from the absence of compression artifacts in our reconstructions, which were present in the compressed images. For two of the salient images, nearly 100\% of respondents favored our reconstructions, showing effective alignment with human perception.

%% file: discussion.tex
\section{Discussion and Future Work}
There are several areas of future exploration that can help achieve a better bandwidth-quality tradeoff while addressing regulatory, privacy, and ethical concerns. These include:

\noindent\textbf{Adaptive Compression Techniques.} Developing systems that dynamically choose between traditional compression and AI-based reconstruction based on image content and network conditions could enhance bandwidth-quality trade-offs. AI could handle high-frequency details and textures while conventional methods preserve large-scale structures and color information.

\noindent\textbf{Multi-modal Conditioning.} Creating models that use broader web page contexts, including text, audio, or video, can inform the image reconstruction process, ensuring images are visually appealing and contextually appropriate.

\noindent\textbf{Client-Side Reconstruction.} With the emergence of lightweight language models designed for smartphones, such as Apple's OpenELM~\cite{openelm}, there is potential for running text-to-image models on these devices in the future. Additionally, the use of the latest one-step generators, which can produce images in \(\frac{1}{50}\)th of the usual inference times~\cite{yin2024improved} shows promise in viable client-side image reconstruction. This can greatly improve internet affordability.

\noindent\textbf{Regulatory and Standardization Efforts.} Collaborating with standards bodies to develop guidelines for AI-based image transmission and reconstruction, addressing legal and regulatory challenges related to image rights and manipulation.

\noindent\textbf{Ethical Considerations.} Addressing privacy concerns in reconstructing images, especially those with personal information, and mitigating misuse in creating or altering images without consent. Considering bias in generative models and its impact on image reconstruction.